\documentclass[twocolumn,amsmath,amssymb]{revtex4-1}

\usepackage[textsize=tiny,backgroundcolor=yellow]{todonotes}

\usepackage{bm}

\newcommand*{\fref}[1]{Fig.~\ref{#1}}
\newcommand{\bra}[1]{\left\langle#1\right|}
\newcommand{\ket}[1]{\left|#1\right\rangle}
\def\cm{cm$^{-1}$}

\begin{document}
    
    \title{Multi-Layer Multi-Configuration Time-Dependent Hartree (ML-MCTDH) Approach to the Correlated Exciton-Vibrational Dynamics in the FMO Complex}
    
    \author{Jan Schulze}
    \affiliation{
        Institut f\"{u}r Physik, Universit\"{a}t Rostock, Albert-Einstein-Str. 23-24, 18059 Rostock, Germany
    }%
        \author{Mohamed F. Shibl}
\thanks{Permanent address: Faculty of Science, Department of Chemistry, Cairo University, Giza, Egypt.}
    \author{Mohammed J. Al-Marri}
    \affiliation{
       Gas Processing Center, College of Engineering, Qatar University, P.O. Box 2713, Doha, Qatar
    }%
    \author{Oliver K\"uhn}
    \email{oliver.kuehn@uni-rostock.de}
    \affiliation{
        Institut f\"ur Physik, Universit\"{a}t Rostock, Albert-Einstein-Str. 23-24, 18059 Rostock, Germany
    }%
    
    \date{\today}
\begin{abstract}
The coupled quantum dynamics of excitonic and vibrational degrees of freedom is investigated for high-dimensional models of the 
Fenna-Matthews-Olson (FMO) complex. This includes a seven and an eight-site model with 518 and 592 harmonic vibrational modes, respectively. The coupling between local electronic transitions and vibrations is described within the Huang-Rhys model using parameters that are obtained by discretization of  an experimental spectral density. Different pathways of excitation energy flow are analyzed in terms of the reduced one-exciton density matrix, focussing on the role of vibrational and vibronic excitation. Distinct features due to both competing time scales of vibrational and exciton motion and vibronically-assisted transfer are observed. The question of the effect of initial state preparation is addressed by comparing the case of an instantaneous Franck-Condon excitation at a single site with that of a laser field excitation.
\end{abstract}
\maketitle

\section{Introduction} 
\label{intro}
The observation of signatures of long-lasting coherent dynamics in photosynthetic antenna complexes~\cite{engel07_782,panitchayangkoon10_12766,collini10_644,panitchayangkoon11_20908} has stimulated theoretical investigations towards unraveling the role of coherence and thus quantum mechanics in photosynthetic light-harvesting (for reviews, see Refs.~\cite{ishizaki10_7319,huelga13_181,schroter15_1}). Here, the coupling between electronic (excitonic) and nuclear degrees of freedom (DOFs) has been in the focus of the discussion. On the one hand side, it causes phase and energy relaxation within the excitonic subsystem and, therefore, is vital for the realization of the directed energy transfer. On the other hand side, the role of exciton-vibrational coupling (EVC) appears to be more intricate, with vibrations being able, e.g., to promote excitation energy transfer~\cite{womick11_1347,chin13_113} or to give rise to specific spectroscopic features \cite{christensson12_7449,chenu13_2029,tiwari13_1203,plenio13_235102,schroter15_1}. Indeed, due to resonance effects excitonic and vibronic excitations can be mixed, despite the smallness of the Huang-Rhys factor~\cite{christensson12_7449,polyutov12_21,schulze14_045010,schroter15_536}.

Exciton transfer in photosynthetic light-harvesting is a problem of dissipative quantum dynamics \cite{may11,renger01_137}. The non-trivial role played by  EVC suggests to apply non-perturbative and non-Markovian approaches such as the quasi-adiabatic path integral \cite{makri98_4144,nalbach15_022706,mujica-martinez15_592} or the hierarchy equation of motion \cite{tanimura06_082001,dijkstra12_073027,strumpfer12_2808,zhu11_1531,kreisbeck12_2828,hein12_023018} method. The basis for these simulations is usually the Frenkel exciton description (the relevant system), combined with a linear Huang-Rhys like coupling of local electronic excitations to vibrational DOFs (the bath)~\cite{schroter15_1}. Alternatively, a dual-bath approach has been proposed. Here, a few selected primary vibrational modes are taken as part of the system, which coupled to the remaining bath modes in a Caldeira-Leggett fashion~\cite{matro95_2568,renger96_15654,kuhn96_99,liu16_}. This provides useful, whenever non-perturbative and non-Markovian effects are important for a few modes only and the remaining bath can be treated in Markov approximation. A combination of this approach with an exact treatment of the bath has been presented in Refs.~\cite{nalbach15_022706,mujica-martinez15_592}, but its computational demand requires to approximate the dynamics of the relevant system. There are two reasons why the latter approach is nevertheless of great interest. First, working with the reduced density operator of the relevant system all explicit information of the bath dynamics is lost, i.e. by construction the interaction is only reflected in system observables such as spectra. Given the prominent role of specific vibrations for the dynamics of light-harvesting proteins this is, of course, a drawback as far as the direct interpretation is concerned. Second, in system-bath approaches the bath is commonly treated at the level of fluctuations with respect to the thermal equilibrium state, which is assumed to be maintained during the dynamics. This so-called linear response limit requires that the bath forms a dense manifold of states, which interacts weakly and essentially uniformly with the system \cite{makri99_2823}. Hence, linear response falls short in describing situations where, e.g., certain bath modes interact rather specifically  with the relevant system, e.g., due to resonance effects. Besides the above mentioned dual bath approaches, which rely on some ad hoc separation of bath modes, there are systematic attempts to tackle the issue of non-equilibrium bath dynamics. For instance, the time-dependent projection operator technique provides a rigorous derivation of coupled equations for non-equilibrium system and bath density operators~\cite{willis74_1343,linden98_411}, although the resulting nonlinear equations have not been tackled in the present context.

With the development of the highly efficient multi-configuration time-dependent Hartree wave packet method~\cite{meyer90_73,beck00_1,meyer03_251,meyer11_351} and in particular its multi-layer extension (ML-MCTDH)~\cite{wang03_1289,manthe08_164116,vendrell11_044135}, it became possible to approach the continuous limit by discretization of the  bath spectral density using thousands of DOFs~\cite{wang00_9948,nest03_24,wang08_115005,bonfanti15_124703}. The combination with imaginary time propagation of the Boltzmann operator using either stochastic thermal wave functions \cite{manthe01_321,nest07_134711,lorenz14_044106} or Monte Carlo importance sampling \cite{wang06_034114} even allows for inclusion of finite temperature effects. As a consequence, ML-MCTDH outperformed the original MCTDH density matrix formulation~\cite{raab99_8759}, which found only a few applications (see, e.g., Ref.~\cite{cattarius04_9283,bruggemann08_152}). 

It should be noted that ML-MCTDH can be understood as being a low-rank tensor decomposition scheme (for a review, see~\cite{grasedyck13_53}). It shares this formal background with the time-dependent density matrix renormalization group approach \cite{prior10_050404}, which has been applied to exciton dynamics of dimer systems in the limit of strong coupling to vibrations \cite{prior10_050404,chin13_113}. Key to that method is a combination with a mapping of the coupled bath modes onto a one-dimensional chain of effective modes with nearest neighbor couplings. This is bound, however, to a linear system-bath coupling, a restriction, which does not exist for the ML-MCTDH method.

Recently, we have applied ML-MCTDH to the problem of the coupled exciton-vibrational dynamics in a model of the FMO complex~\cite{schulze15_6211}. Thereby, we have restricted ourselves to the description of three sites only, each being coupled to 150 vibrational DOFs. Coupling parameters (Huang-Rhys factors) and frequencies of the modes were obtained by discretization of an experimental spectral density~\cite{wendling00_5825} up to 300 \cm. The dynamics was followed after instantaneous Franck-Condon excitation of the first site. It turned out that under these conditions, in particular, modes in the range between 160 and 300 \cm{} are responsible for the subpicosecond decay of excitonic populations and coherences. Effects of vibronic resonance-assisted exciton transfer have been observed for modes around 180 \cm. Further, there has been an appreciable vibrational excitation in the electronic ground states of those sites that are not electronically excited.

In the present contribution, the study of Ref.~\cite{schulze15_6211} is extended in several respects. First, full seven and eight sites models of the FMO complex are considered. Besides numerical feasibility, we will focus on the question to what extent alternative pathways from the initially excited to the sink site are taken. Second, the issue of correlation is addressed. In the three-site model the Hartree approximation badly failed in reproducing the full ML-MCDTH dynamics~\cite{schulze15_6211}. Here, we inspect the performance of this approximation in the light of the different excitonic pathways of the full model. Third, the effect of explicit excitation with a laser field on the exciton-vibrational dynamics is investigated.

The paper is organized as follows: In Section \ref{sec:methods} the model Hamiltonian is defined and a brief introduction into the ML-MCTDH method is provided. Section \ref{sec:results} starts with the field-free dynamics of the seven- and eight-site models. Next the effect of preparation by a laser field with finite duration is addressed. The paper is summarized in Section \ref{sec:concl}.
\section{Theoretical Methods}
\label{sec:methods}
\subsection{Exciton-Vibrational Hamiltonian} 
The Frenkel exciton Hamiltonian describes an aggregate with $ N_{\rm agg} $ sites (site index $m$), each site having the excitation energy $E_m$, and different sites being coupled by the Coulomb interaction $J_{mn}$\cite{may11}
\begin{equation} 
H_{\rm ex} =\sum_{m,n=1}^{N_{\rm agg}}(\delta_{mn} E_m + J_{mn})\ket{m}\bra{n} \, .
\end{equation}
Local electronic states are restricted to the ground $\ket{g_m}$ and excited states $\ket{e_m}$, i.e. the Frenkel zero- and one-exciton states are given by $\ket{0}=\prod_{m}\ket{g_m}$ and $\ket{m}=\ket{e_m}\prod_{n\ne m}\ket{g_n}$, respectively. For the present simulations, we will use the eight-site FMO Hamiltonian reported by Moix et al. which is given here for completeness  (in units of \cm, off-set is 12195 \cm)~\cite{moix11_3045}:
\begin{eqnarray}
\label{eq:hmat}
\mathbf{H}_{\mathrm{ex}}&=&\left(  
    \begin{array}{cccccccc}
      310&{-98}&6&-6&7&-12&-10&{38}\\
      {-98}&230&{30}&7&2&12&5&8\\
      6&{30}&0&{-59}&-2&-10&5&2\\
      -6&7&{-59}&180&{-65}&-17&{-65}&-2\\
      7&2&-2&{-65}&405&{89}&-6&5\\
      -12&11&-10&-17&{89}&320&{32}&-10\\
      -10&5&5&{-64}&-6&{32}&270&-11\\
	{38}&8&2&-2&5&-10&-11&505
    \end{array}                    
  \right)\,.\nonumber\\
  &&
\end{eqnarray}
It is a combination of  site energies obtained from quantum chemical/electrostatic calculations~\cite{schmidtambusch11_93} and Coulomb couplings described within the dipole-dipole approximation~\cite{moix11_3045}. The labeling of the sites follows the structure of the Hamiltonian matrix, e.g., site $m=3$ is the energetically lowest site, which is connected to the cytoplasmic membrane containing the  reaction center complex, and site $m=8$ is the highest in energy and believed to act as a linker between the baseplate and the FMO complex.

\begin{figure}[t]
\includegraphics[width=0.5\textwidth]{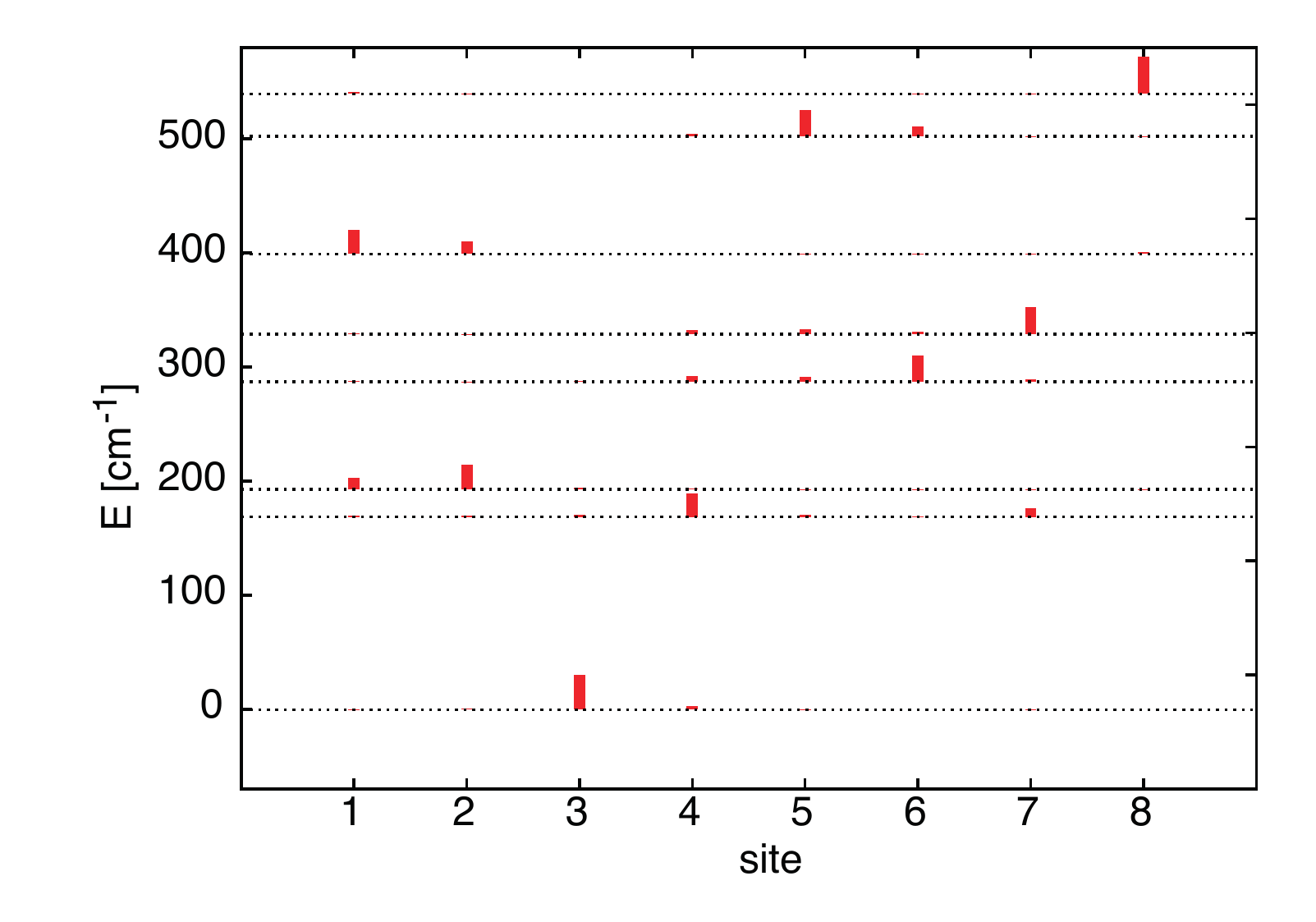}
\caption{
(color online) Spectrum of one-exciton eigenstates and their decomposition into local states (red bars give the squared local amplitudes 
$|c_{m}(\alpha)|^2$).
}
\label{fig:levels}     
\end{figure}

Diagonalization of this matrix yields the one-exciton eigenstates $|\alpha \rangle = \sum_m c_{m}(\alpha) |m\rangle$,
 whose energies and decompositions into the local states $\ket{m}$ are shown in Fig.~\ref{fig:levels}.

The local vibrations at site $m$ are described in harmonic approximation by the set of dimensionless normal mode coordinates $ \{ Q_{m,\xi} \} $ with frequencies $ \{ \omega_{m,\xi} \} $, i.e. the vibrational Hamiltonian reads
\begin{equation}
H_{\rm vib} =\sum_m \sum_{\xi \in m}\frac{\hbar\omega_{m,\xi}}{2} \left( - \frac{\partial^2}{\partial Q_{m,\xi}^2}+ Q_{m,\xi}^2\right) \, .
\end{equation}
EVC is accounted for within the linearly shifted oscillator (Huang-Rhys) model, i.e.
\begin{equation}
H_{\rm ex-vib} = \sum_m \sum_{\xi \in m} \hbar \omega_{m,\xi} \sqrt{2 S_{m,\xi}}Q_{m,\xi} \ket{m}\bra{m}\, .
\end{equation}
The coupling of a particular mode to the electronic transition is characterized by the Huang-Rhys factor $S_{m,\xi}$.

The initial excitation by an external laser field is realized within the dipole-approximation (for the MCTDH implementation, see Ref.~\cite{naundorf02_719})
\begin{equation}
\label{eq:hfield}
	H_{\rm f}(t) = -{\mathcal E}(t) \sum_m d_m |m\rangle\langle 0 | + {\rm h.c.}\,.
\end{equation}
Since we are only interested in the difference between instantaneous and finite field excitation without aiming at a comparison with experiment,  the effect of different orientations of the chromophore with respect to the field polarization is neglected. For the laser field we will assume a Gaussian envelope
\begin{equation}
	{\mathcal E}(t) = {\mathcal E}_0 \cos(\Omega t) \exp[-2\ln(2)	 (t-t_0)^2/\tau^2] \, .
\end{equation}
Here, ${\mathcal E}_0$ is the field amplitude, $\Omega$ is the carrier frequency, $\tau$ is the full pulse width at half maximum (FWHM), and $t_0$ is the pulse center.

\begin{figure}[t]
\begin{center}
\includegraphics[width=0.5\textwidth]{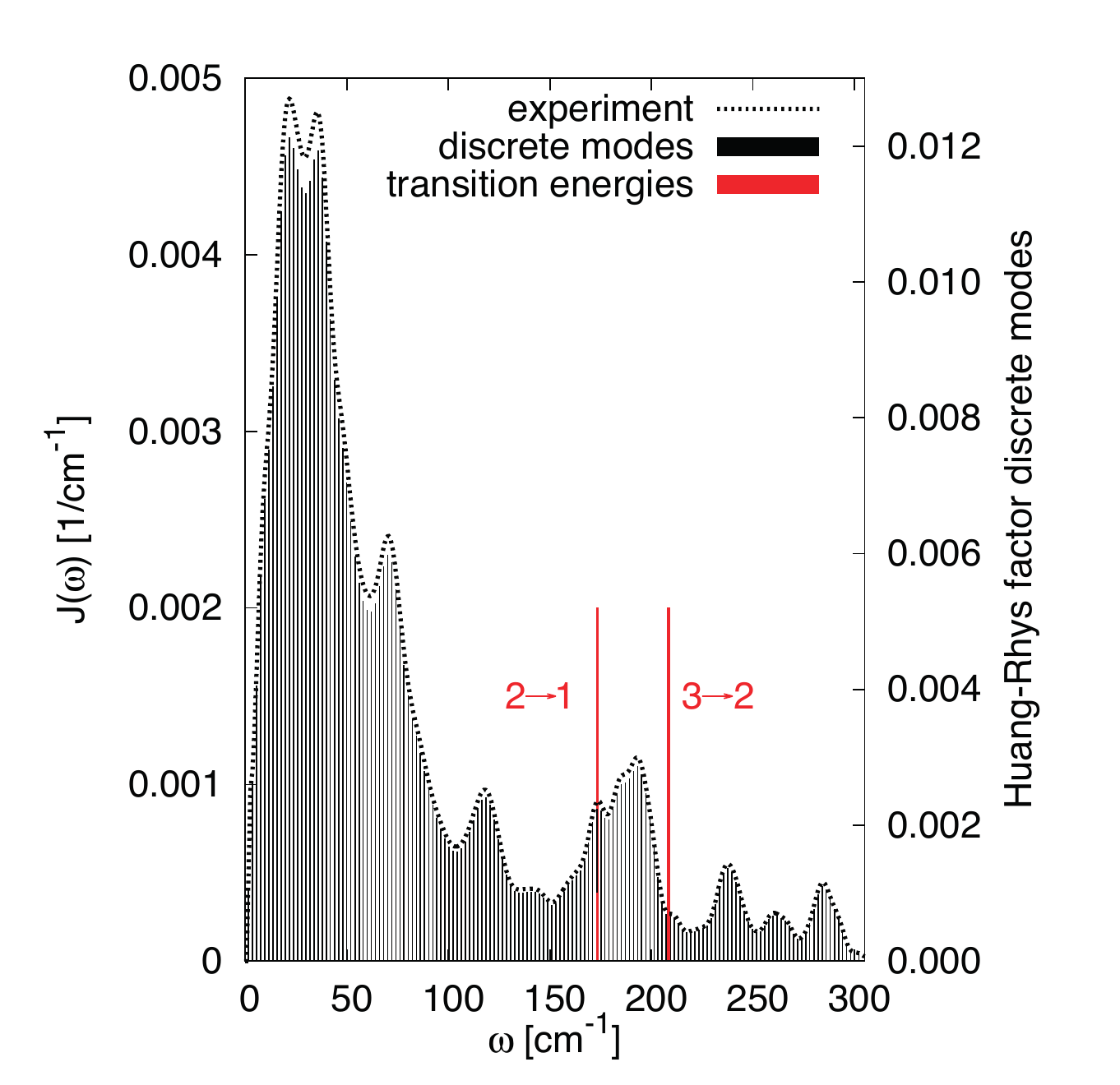}
\caption{(color online) Experimental spectral density for an FMO  BChl $a$ molecule~\cite{wendling00_5825}  and approximate stick spectrum used in the present 74 mode model. Also shown are the transition energies for exciton eigenstates as indicated.}
\label{fig:specdens}     
\end{center}
\end{figure}

Vibrational excitation in the electronic ground and excited state will be called vibrational and vibronic excitation, respectively. The energy of the vibrational excitation at site $m$ can be obtained from the expectation value of the Hamiltonian operator
\begin{equation}
\label{eq:vibra}
H^{\rm (vibra)}_m=\sum_{\xi \in m}\frac{\omega_{m,\xi}}{2}\left(-\frac{\partial^2}{\partial Q_{m,\xi}^2}+Q_{m,\xi}^2\right)(1-|m\rangle\langle m|) \,.
\end{equation}
Note that this expression gives the vibrational energy irrespective which site of the aggregate is electronically excited.
The vibronic energy at site $m$ is defined by the Hamiltonian
\begin{eqnarray}
\label{eq:vibro}
H^{\rm (vibro)}_m &=&
\sum_{\xi \in m} \frac{\omega_{m,\xi}}{2}\left(-\frac{\partial^2}{\partial Q_{m,\xi}^2}+Q_{m,\xi}^2+ 2 \sqrt{2 S_{m,\xi}}  Q_{m,\xi} \right)\nonumber\\ &\times & |m\rangle\langle m|  \,.
\end{eqnarray}
Frequencies and Huang-Rhys factors can be obtained from the spectral density, $J_m(\omega)$, of the monomeric BChl~$a$ molecule~\cite{may11}
\begin{equation}
\label{eq:jomega}
	J_m(\omega) = A \sum_{\xi\in m}   S_{m,\xi} \delta(\omega-\omega_{m,\xi})\, ,
\end{equation}
where $A$ is a constant that will be used to adjust the total HR factor for site $m$ for a finite discretization according to $S_{\rm tot}=A^{-1}\int d\omega J_{m}(\omega) = \sum_{\xi \in m} S_{m,\xi}$.
There are several simulations of the spectral density, taking into account the protein and solvent environment~\cite{olbrich11_1771,renger12_14565,rivera13_5510,valleau12_224103}. 
 Since the reported results differ considerably we will use the experimentally determined spectral density of Wendling et al.~\cite{wendling00_5825} shown in \fref{fig:specdens}. It has been obtained from low-temperature site-selected fluorescence, measured for the energetically lowest pigment of the complex. The total HR factor was determined as $S_{\rm tot}=0.42$. Note that the experimental spectral density covers the range up to about 350 \cm{} only. Since excitonic transition frequencies between strongly coupled pigments are essentially located in the range up to 300 \cm{} the neglect of  higher frequency modes is justified.

 In the present model  we will use the original spectral density from Ref.~\cite{wendling00_5825} and discretize it into 74 modes within the interval $[2:300]$ \cm.  The amplitudes of the individual HR factors have been adjusted homogeneously via the constant $A$ such as to  preserve $S_{\rm tot}=0.42$ upon summation. Notice that in our previous work~\cite{schulze15_6211} a discretization into 150 modes had been used. Both discretizations formally yield  recurrence times, $T_{\rm rec}=2\pi/\Delta \omega$, that are well beyond the time scale considered here (1 ps). In fact test calculations using the previous three-site model gave no noticeable difference in the population dynamics.
 
 Notice that there are alternative ways to treat the experimental spectral density. For instance, it could be separated into a structureless phonon wing plus a part, which takes into account the discernible broadened peaks, e.g., in the spirit of a multi-mode Brownian oscillator model (see also Ref. \cite{adolphs06_2778} where the structured part of the spectral density is described by a single effective discrete vibration at 180 \cm). Still another choice would be a fit of the total spectral density, e.g., to a superposition of simple Drude-Lorentz functions \cite{kreisbeck12_2828}. The present treatment avoids such decompositions and treats the whole spectral density on the same footing.
%
%
\begin{figure*}[tb]
\begin{center}
\includegraphics[width=0.95\textwidth]{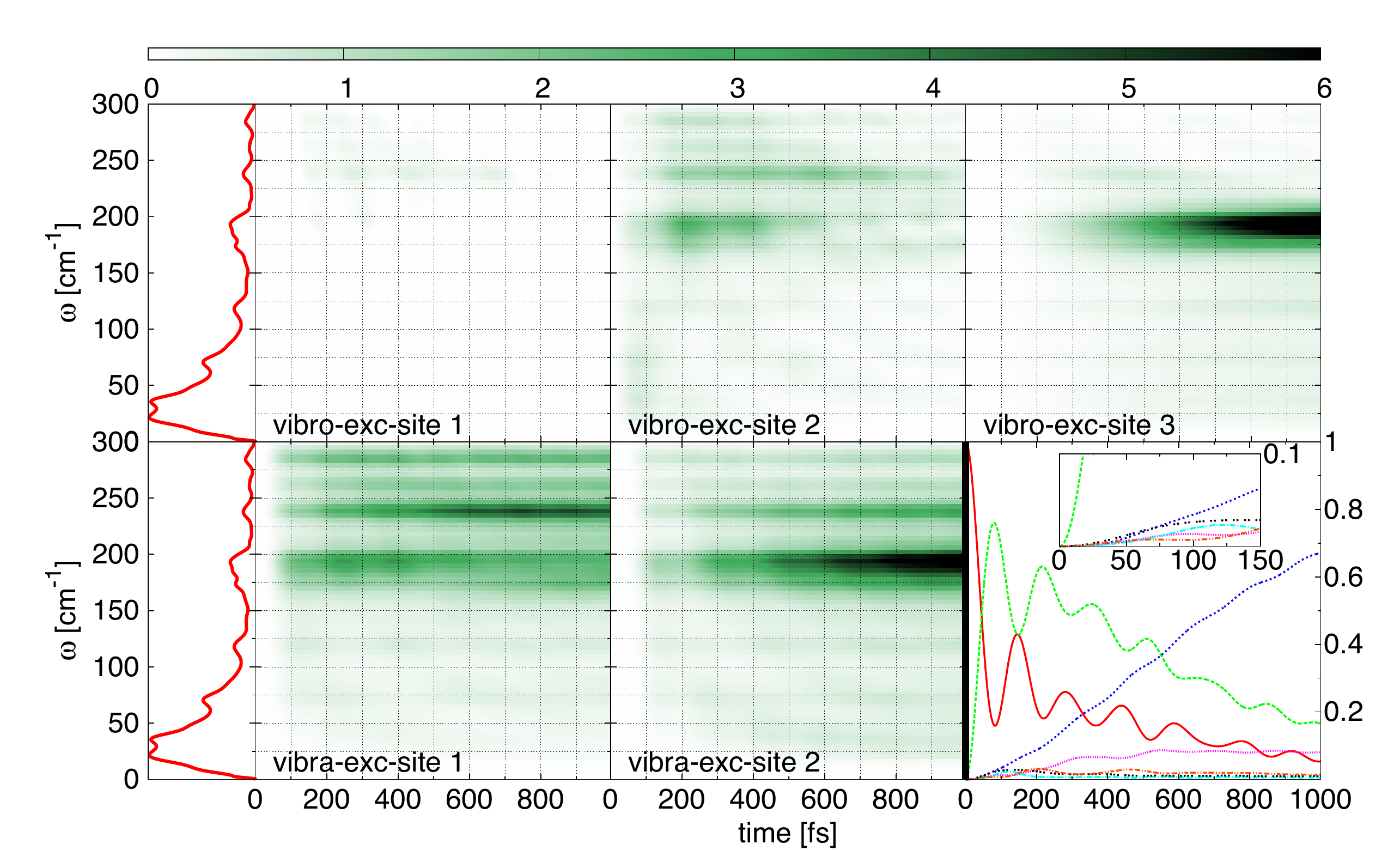}
\caption{(color online) Population dynamics (lower right) of the seven-site FMO model after instantaneous excitation of site $m=1$. The other panels show the vibrational (lower row) and vibronic (upper row) energy for selected sites and for  each mode according to Eqs.~(\ref{eq:vibra}) and (\ref{eq:vibro}), respectively. In the left part the spectral density is shown, cf. Fig.~\ref{fig:specdens}. The color key for the site populations is given in Fig.~\ref{fig:7_vib}.} 
\label{fig:7_pop}     
\end{center}
\end{figure*}
%

%
\begin{figure}[tb]
\begin{center}
\includegraphics[width=0.5\textwidth]{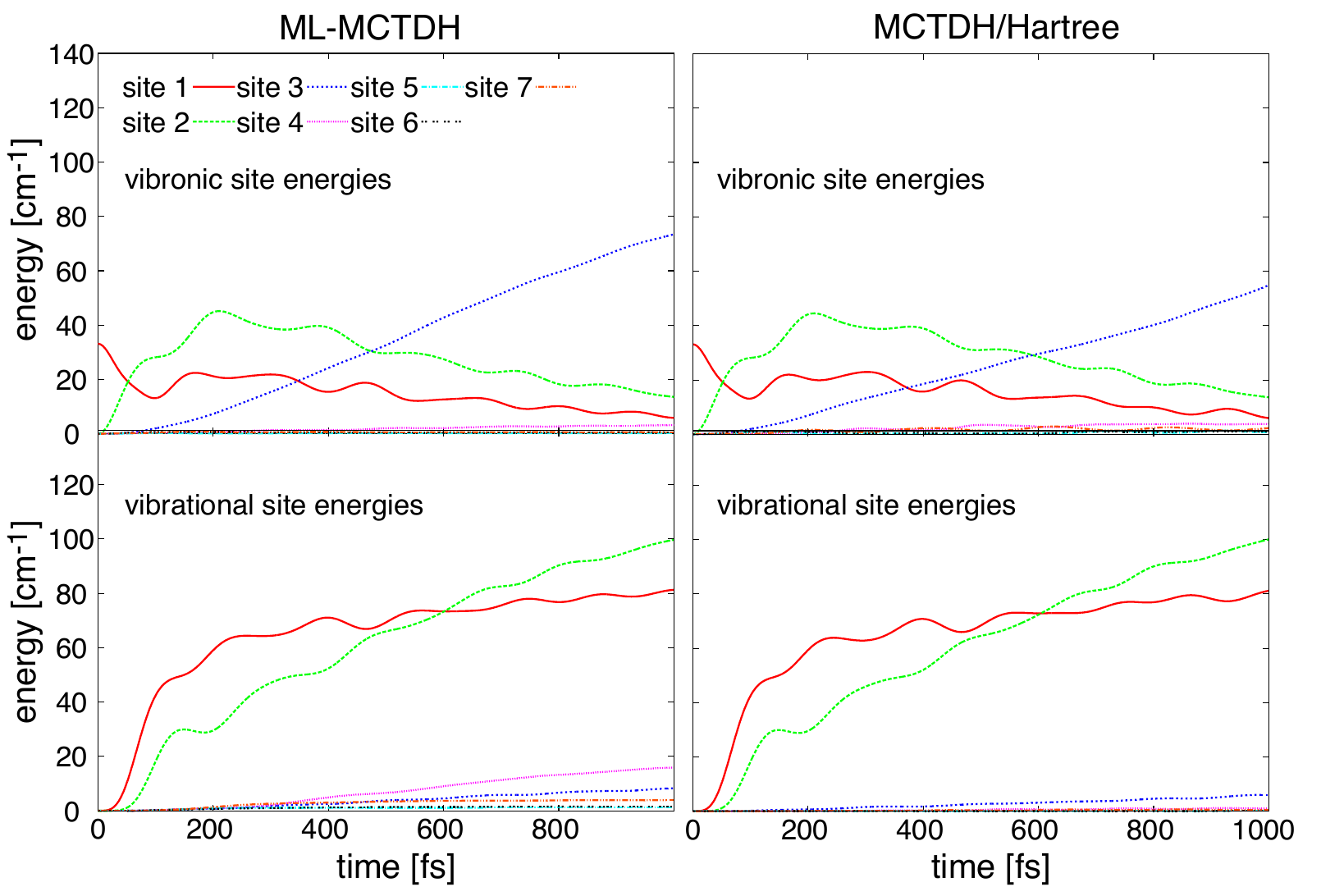}
\caption{(color online) Energy expectation values for the vibrational and vibronic Hamiltonian according  to Eqs.~(\ref{eq:vibra}) and (\ref{eq:vibro}), respectively. Left column: seven-site ML-MCTDH model, right column: seven-site MCTDH/Hartree model. }
\label{fig:7_vib}     
\end{center}
\end{figure}
%

%
\begin{figure}[tb]
\begin{center}
\includegraphics[width=0.4\textwidth]{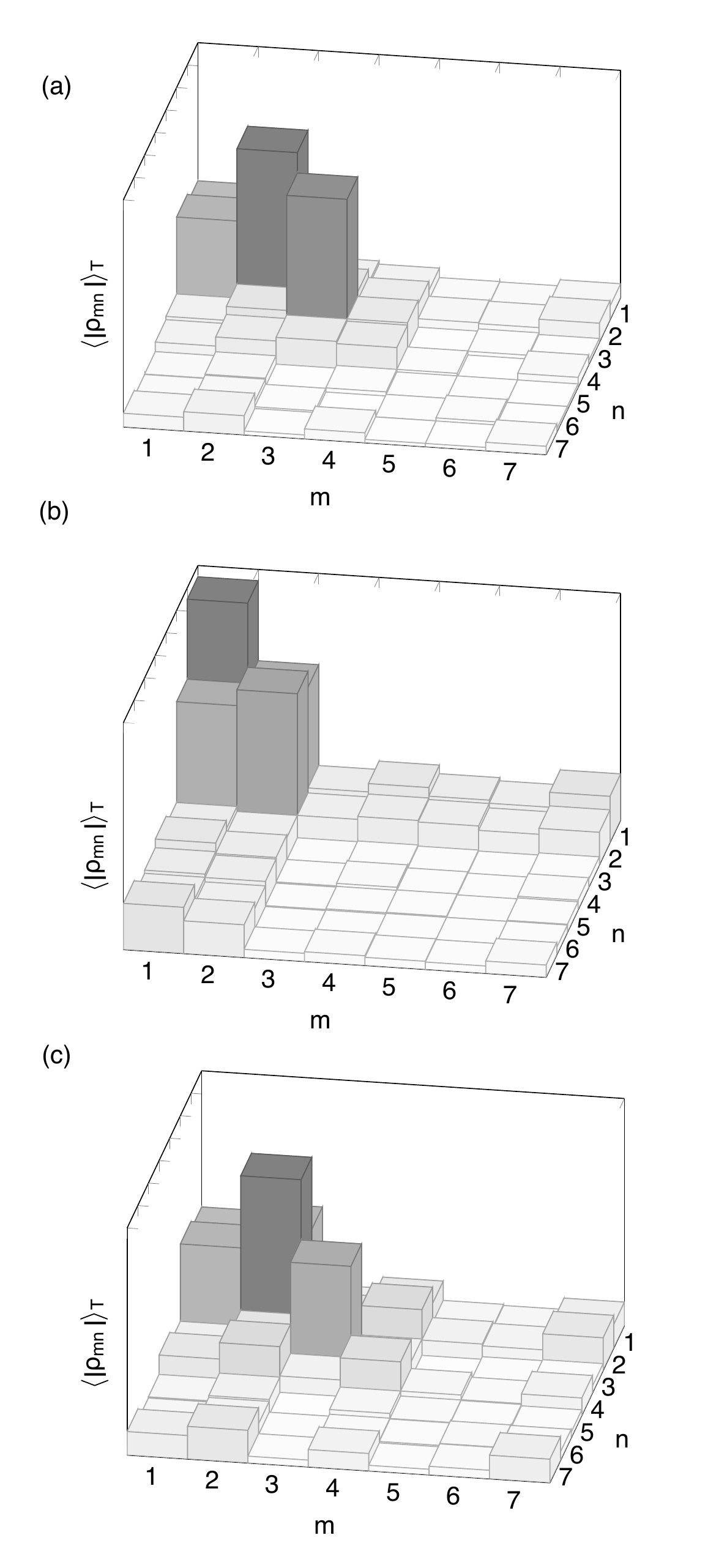}
\caption{Time-averaged density matrix, Eq.~(\ref{eq:av_rdm}) for the seven-site FMO model (a), the bare electronic model (b), and the MCTDH/Hartree model (c).}
\label{fig:7_avrdm}     
\end{center}
\end{figure}
%

%
\begin{figure}[t]
\begin{center}
\includegraphics[width=0.5\textwidth]{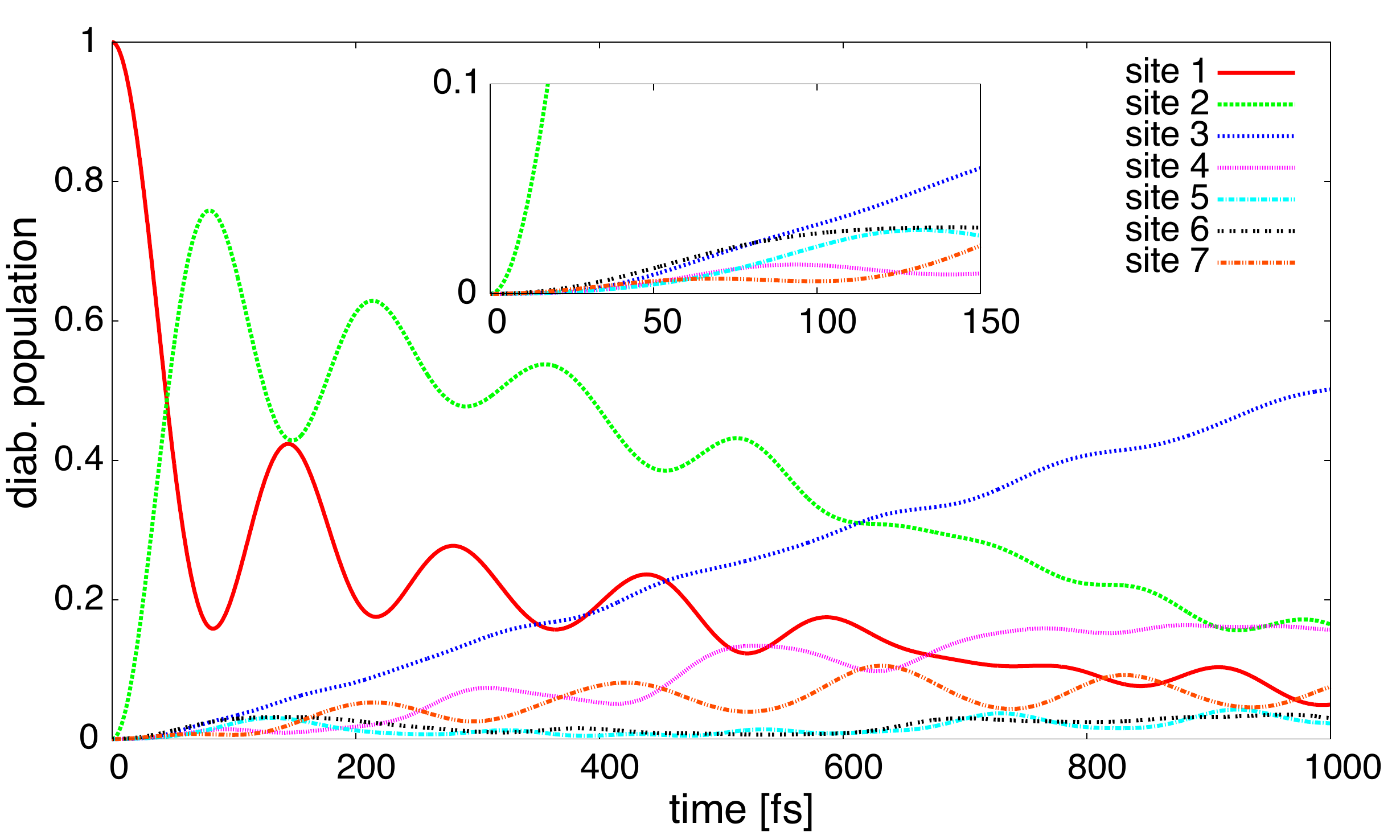}
\caption{(color online) Population dynamics of the seven-site FMO model for the case, where the vibrational DOFs at sites $m=4-7$ are described in Hartree approximation (MCTDH/Hartree).}
\label{fig:7_hart}     
\end{center}
\end{figure}
%

\begin{figure*}[bt]
\begin{center}
\includegraphics[width=0.95\textwidth]{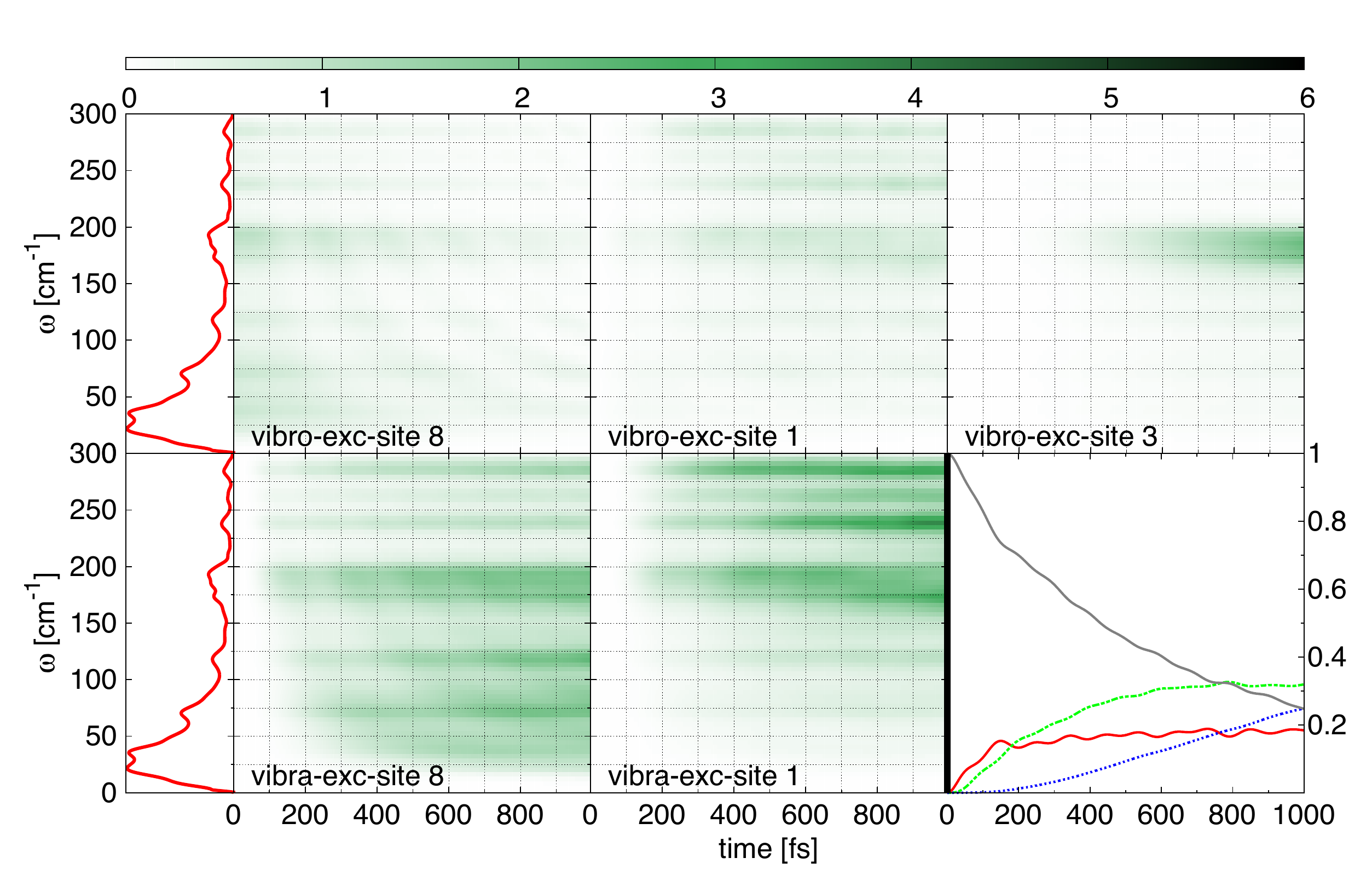}
\caption{(color online) Population dynamics (lower right) of the reduced eight-site FMO model (8123) after instantaneous excitation of site $m=8$. The other panels show the vibrational (lower row) and vibronic (upper row) energy for selected sites and for each mode according to Eqs.~(\ref{eq:vibra}) and (\ref{eq:vibro}), respectively. In the left part the spectral density is shown, cf.\ Fig.~\ref{fig:specdens}. The color key for the site populations is given in Fig.~\ref{fig:7_vib}.}
\label{fig:8_pop}     
\end{center}
\end{figure*}

%
\begin{figure}[tb]
\begin{center}
\includegraphics[width=0.5\textwidth]{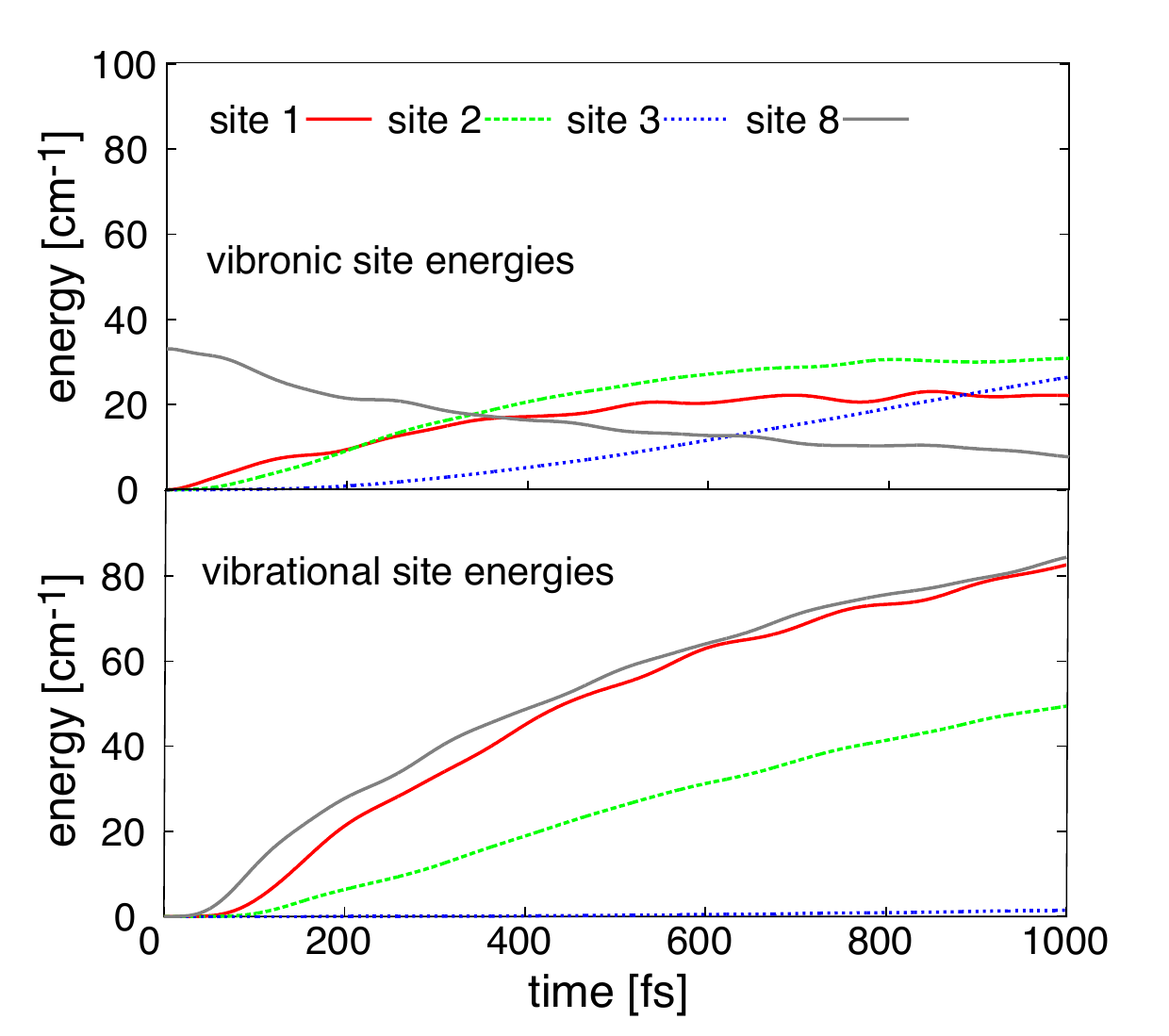}
\caption{(color online) Energy expectation values of the reduced eight-site model (8123) for the vibrational (lower panel) and vibronic (upper panel) Hamiltonian according  to Eqs.~(\ref{eq:vibra}) and (\ref{eq:vibro}), respectively.}
\label{fig:8_vib}     
\end{center}
\end{figure}

%
\begin{figure}[tb]
\begin{center}
\includegraphics[width=0.5\textwidth]{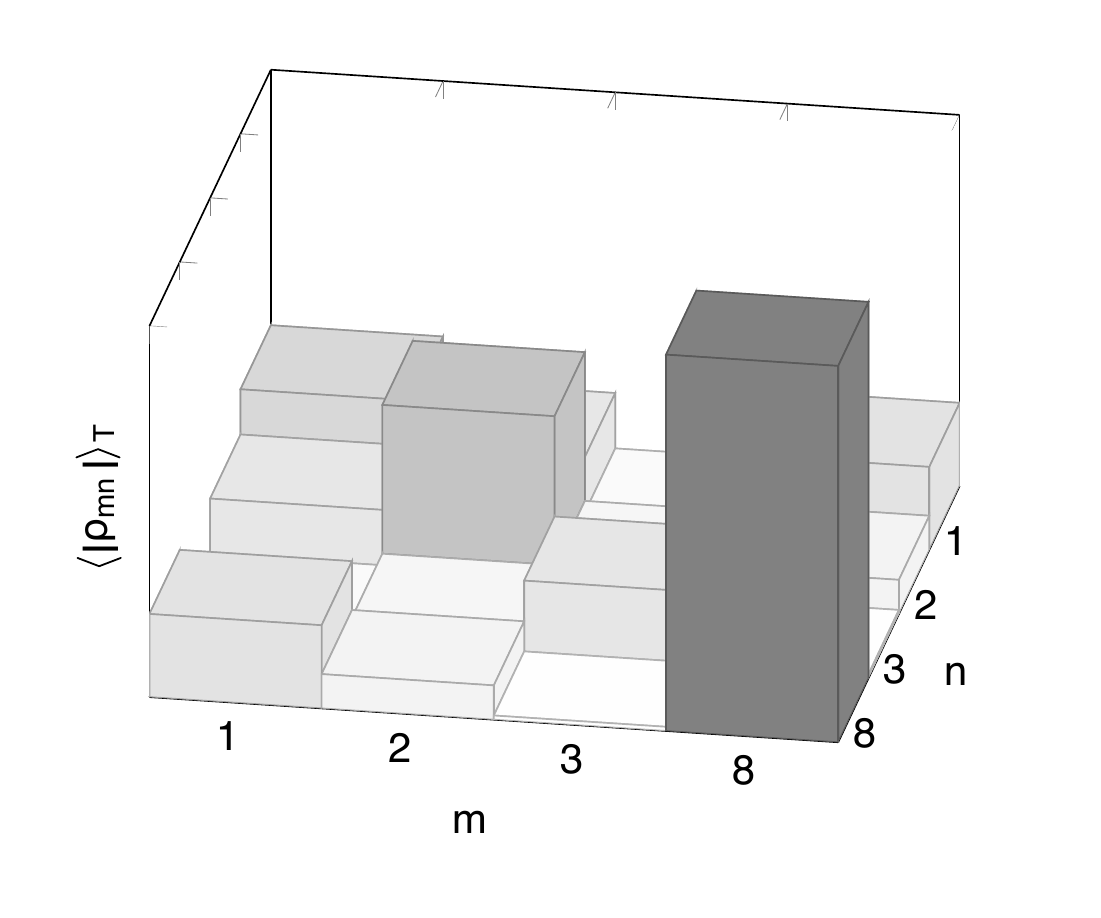}
\caption{ Time-averaged density matrix, Eq.~(\ref{eq:av_rdm}) for the reduced eight-site FMO model (8123).}
\label{fig:8_avrdm}     
\end{center}
\end{figure}

%
%
\begin{figure*}[bt]
\begin{center}
\includegraphics[width=0.85\textwidth]{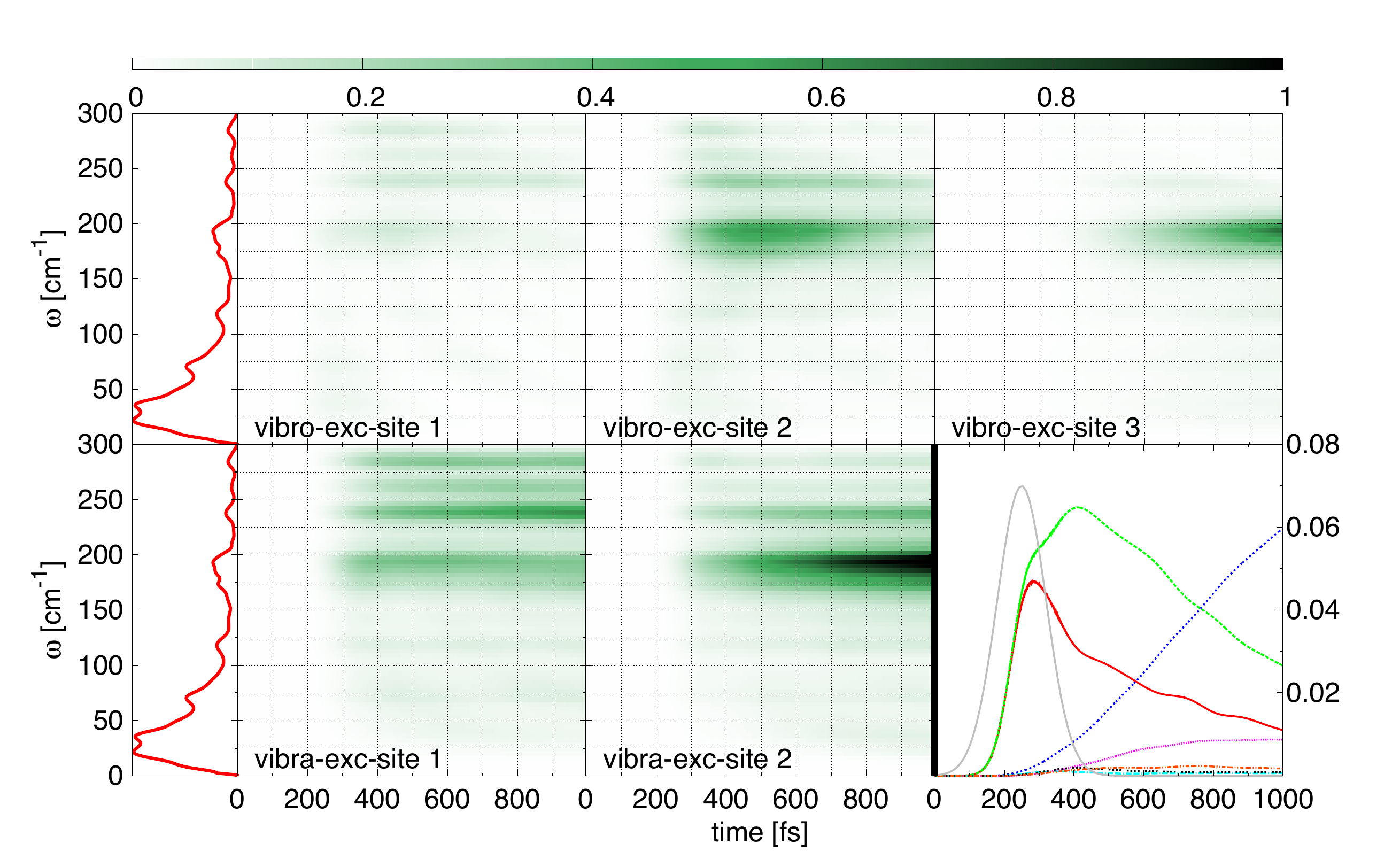}
\caption{(color online) Population dynamics (lower right) of the seven-site FMO model after  excitation of sites $m=1$ and 2 with a laser field. The grey curve represents the field envelope. Field parameters are: ${\mathcal E}_0=0.3$~mE$_h$/ea$_B$, $\Omega=12574$~\cm, $\tau=99$fs$^{-1}$, and $t_0=250$~fs. The other panels show the vibrational (lower row) and vibronic (upper row) energy for selected sites and for  each mode according to Eqs.~(\ref{eq:vibra}) and (\ref{eq:vibro}), respectively. In the left part the spectral density is shown, cf.\ Fig.~\ref{fig:specdens}. The color key for the site populations is given in Fig.~\ref{fig:7_vib}.  }
\label{fig:7_pop_field}     
\end{center}
\end{figure*}
%

\subsection{Quantum Dynamics}
The time-dependent Schr\"odinger equation will be solved employing the ML-MCTDH method (for a review, see Ref.~\cite{meyer11_351}). The state vector is expanded into the local exciton basis according to
\begin{eqnarray}
	|\Psi({\bf Q};t) \rangle=\sum_{\kappa} \chi_{\kappa }({\bf Q};t) \, |\kappa \rangle	\, \quad \quad \kappa \in (0, m) \,.
\end{eqnarray}
The  nuclear coordinates are comprised into the $D=N_{\rm agg}\times N_{\rm vib}$ dimensional vector $\mathbf{ Q}$.  Here, $N_{\rm vib}$ is the number of modes per site, which is assumed to be site-independent. The nuclear wave function is  expanded into MCTDH form 
\begin{equation}
\label{eq:psiMCTDH}
\chi_\kappa(\mathbf{ Q},t) = \sum_{j_1 \ldots j_D}^{{n_{j_1} \ldots n_{j_D}}}
C^{(\kappa)}_{j_1,\ldots,j_D}(t) \phi^{(\kappa)}_{j_1}(Q_1;t) \ldots \phi^{(\kappa)}_{j_D}(Q_{D};t) \, .
\end{equation}
Here, the $C^{(\kappa)}_{j_1,\ldots,j_D}(t)$ are the time-dependent expansion coefficients weighting the contributions of the different Hartree products, which are composed of $n_{j_{k}}$ single particle functions (SPFs), $\phi^{(\alpha)}_{j_k}(Q_k;t)$, for the $k$th degree of freedom in state $\kappa$. For the zero-exciton state ($\kappa =0$) the nuclear wave function can be written in terms of a single Hartree product since by construction there are no correlations in the exciton ground state. 

In ML-MCTDH the SPFs themselves describe multi-dimensional coordinates that are expanded into MCTDH form~\cite{wang03_1289,manthe08_164116,vendrell11_044135}. This yields a nested set of expansions that can be represented by so-called ML-MCTDH trees~\cite{manthe08_164116}.  The particular choice of this tree  strongly influences the required numerical effort~\cite{vendrell11_044135,schroter15_1}; for applications to coupled electron-vibrational dynamics, see also Refs.~\cite{meng12_134302,meng13_014313}. In the following simulations we use a grouping according to the magnitude of the HR factor as detailed in the Supplementary Material\cite{suppl}, see also Ref.~\cite{schulze15_6211}. 

Wave packet propagations have been performed using the Heidelberg program package \cite{mctdh85}. Temperature effects due to the thermal population of vibrational states in the electronic ground state are not included. In the field-free cases the initial conditions has been a vertical Franck-Condon transition at site $m=1$ (seven-site model) or at site $m=8$ (eight-site model)  and the propagation time was 1 ps. Convergence of the ML-MCTDH setup has been monitored by means of the grid size, the precision of the integrator, and the  natural orbital populations \cite{beck00_1}. The largest  population of the least occupied natural orbital was typically $\sim 10^{-4}$.

The quantum dynamics will be characterized by means of the  one-exciton density matrix
\begin{equation}
\label{eq:av_rdm}
	\rho_{mn}(t) = \langle m |\Psi(t) \rangle \langle \Psi(t) | n \rangle\, .
\end{equation}
Since this expression implies tracing out the vibrational DOFs, $\rho_{mn}(t)$ is actually a reduced density matrix for the exciton subsystem. In order to quantify the contribution of different density matrix elements, their averages with respect to the considered time interval ($T=1$ ps) will be considered~\cite{kuhn97_809}
\begin{equation}
	\langle |\rho_{mn}| \rangle_T = \frac{1}{T} \int_0^T dt\, |\rho_{mn}(t)| \, .
\end{equation}
%

\section{Results}
\label{sec:results}
\subsection{Field-free Dynamics}
\subsubsection{Seven-Site Model}
In Fig.~\ref{fig:7_pop}, results for the dynamics of the seven-site FMO model after initial instantaneous excitation of site $m=1$ are given. First, let us focus on the exciton populations, $\rho_{mm}(t)$,  shown in the lower right panel. Apparently, there is a coherent population exchange between the initially occupied site $m=1$ and site $m=2$. The population oscillation has a period of about 150 fs, which is slightly different from the bare electronic case (160 fs) due to EVC. The decay of the populations of these two sites is accompanied by an increase of the population of site $m=3$, i.e. the site which is connected to the reaction center. Notice that there is almost no oscillation of $\rho_{33}$, due to the large energy gap; cf. Fig.~\ref{fig:levels}. Compared to the previous three-site model \cite{schulze15_6211} there are small differences in the population dynamics (cf. Fig.~S2 in the Suppl. Mat.\cite{suppl}). After 1 ps the populations of the seven/three-site model are 0.05/0.08,0.17/0.25, and 0.67/0.66.
This expresses the fact that there is an additional pathway, which involves site $m=4$ as the doorway to the terminal site $m=3$. This yields a slight increase of the population of sites $m=3$ after 1 ps, i.e. it enhances the efficiency of energy transfer through the complex. 

Concerning the distribution of energy into modes of vibrational and vibronic excitations in Fig.~\ref{fig:7_pop}, there is  only a small difference between the full and the reduced model. Vibrational excitation dominates at sites $m=1$ and 2 for those modes whose oscillation period is faster than the inter-site coupling (i.e. with frequencies above 160 \cm). This dynamical effect can be attributed to the competition between transfer and wave packet motion out of the initial Franck-Condon window (see, Ref.~\cite{schulze15_6211}). At site $m=3$ vibronic excitation of modes  around 190 \cm{} is observed, which gives indication for vibronically enhanced exciton transfer (compare the energy gap between sites $m=2$ and 3 in Fig.~\ref{fig:levels}).

A global view on vibrational and vibronic excitation can be obtained from the site-resolved expectation values defined in Eqs.~(\ref{eq:vibra}) and (\ref{eq:vibro}), respectively. The results are given in Fig.~\ref{fig:7_vib}. Here, we notice that vibronic excitations are restricted almost exclusively to sites $m=1-3$, whereas vibrational excitations are observed, apart from small contributions at sites $m=3$ and 4,  mostly for sites $m=1$ and 2. Within the above-mentioned dynamical picture, only the long-lasting coherent population oscillations between these two sites generate an appreciable vibrational excitation.

The analysis of the exciton density matrix facilitates a quantification of inter-site coherences. In Fig.~\ref{fig:7_avrdm}a the time-averaged density matrix according to Eq.~(\ref{eq:av_rdm}) is shown for the full seven-site model (the time-dependence of selected coherences can be found in Fig.~S5 in the Suppl. Mat.\cite{suppl}). As expected coherences and population for sites $m=1$ and 2 are sizable, thus demonstrating the coherent nature of the population oscillations. In addition, the average population of site 3, $\rho_{33}$, is rather large (cf. Fig.~\ref{fig:7_pop}). Further, there is noticeable amplitude at site $m=4$, i.e. for $\rho_{44}$, but also for the coherences $\rho_{43}$, $\rho_{42}$, and $\rho_{32}$. Interestingly, site $m=7$ is connected via coherences to sites $m=1$, 2, and 4. This is not apparent from the pure electronic level scheme in Fig.~\ref{fig:levels}. Indeed, EVC substantially influences the exciton density matrix as can be seen by comparing panels (a) and (b) of Fig.~\ref{fig:7_avrdm}. the latter shows the time-averaged density matrix for a bare electronic model, which does not give any appreciable transfer to site $m=3$ (see also Ref.~\cite{schulze15_6211}). Based on these results, one might argue that EVC dynamics  imprints specific exciton transfer pathways onto the model.

Finally, we focus on a limiting case of the multidimensional wave packet expansion, Eq.~(\ref{eq:psiMCTDH}), which could lead to a drastic reduction of the computational effort, i.e. the Hartree approximation. In Ref.~\cite{schulze15_6211} we had shown, for the three-site model, that the Hartree approximation for the nuclear wave packet of a given exciton state, does not provide a reliable description. However, in the seven-site model most of the dynamics takes place in sites $m=1-3$, i.e. one might argue that correlations beyond the Hartree approximation are important for theses sites only. As a test, we used a setup where only sites $m=1-3$ are described by an MCTDH ansatz, whereas for the other sites a single Hartree product is used (MCTDH/Hartree). In Fig.~\ref{fig:7_hart}, the respective population dynamics is shown. Overall the MCTDH/Hartree model provides a fairly reasonable description of the dynamics of sites $m=1$ and 2, but gives a lower final population of site $m=3$ (0.5).

The distribution of vibrational and vibronic energy expectation values is also rather similar to the full MCTDH case (see, Fig.~S3 in the Suppl. Mat.\cite{suppl}). The smaller value of the final site $m=3$ population can be traced to the erroneous behavior of the dynamics at sites $m=4-7$. Here, one finds a pronounced population oscillation between sites $m=4$ and 7, which reflects the fact that the Hartree ansatz does not provide sufficient flexibility for the wave packet to suppress electronic oscillations in favor of vibrational energy redistribution. This conclusion is supported by the calculated time-averaged density matrix in Fig.~\ref{fig:7_avrdm}c. Compared to the full MCTDH case in panel (a), the magnitude of the off-diagonal elements, like $\rho_{71}$, $\rho_{72}$, and $\rho_{74}$, is too large.
\subsubsection{Eight-Site Model}
The coherent  oscillations of the site populations seen in Fig.~\ref{fig:7_pop} are specific to the preparation of the system at site $m=1$. Moix et al.~\cite{moix11_3045} have discussed an eight-site model, which is likely to be realized in the FMO trimer. Specifically, they have observed that initial preparation of the site $m=8$ does not lead to any population oscillations. This finding has been attributed to the large energy gap between site $m=8$ and the strongest coupled site $m=1$, compare Eq.~(\ref{eq:hmat}) and Fig.~\ref{fig:levels}. In the following we discuss the dynamics of the eight-site model from the perspective of EVC. First, we compared the full eight-site model with a reduced one, which includes the major pathway $8\rightarrow 1 \rightarrow 2 \rightarrow 3$ (8123). Similar to the case of the seven-site model the contribution of the pathway that involves site $m=4$ is rather small and the vibrational and vibronic distributions do not differ much between the eight-site model and its reduced form (see, Figs.~S4 in Suppl. Mat.\cite{suppl}). Therefore, we discuss only the reduced eight-site model (8123) in the following. 

In Fig.~\ref{fig:8_pop}, the population dynamics and the mode-resolved vibrational and vibronic energies are given; cf.\ Fig.~\ref{fig:7_pop}. From the population dynamics, we notice that indeed there are only small periodic modulations of the population of sites $m=8$ and $m=1$. There is no coherent population exchange between sites $m=1$ and $m=2$.  The pattern of vibronic excitations does not look much different if compared to the seven-site model. Since this excitation is a consequence of vibronically-assisted transfer this is no surprise since the gap between the exciton eigenstates dominated by the sites $m=8$ and $m=1$ is about 150 \cm. Interestingly, the vibrational excitation in the electronic ground state of site $m=8$ differs from that of the initial site in the seven-site simulation. In the present case, the extent of excitation resembles the spectral density, i.e.\ low-frequency modes are substantially excited. In view of the dynamical picture discussed above, this is a consequence of the now much longer time scale for transfer as compared to the vibrational period. Notice that modes below 20 \cm{} are still not appreciably excited.

An overview of the total vibrational and vibronic excitation at the different sites is given in Fig.~\ref{fig:8_vib}. Similar to the case of the seven-site model, Fig.~\ref{fig:7_vib}, there is almost no vibrational excitation at site $m=3$, whereas vibronic excitation decreases at the initial site $m=8$ and increases at the final site $m=3$.
Finally, we show the time-averaged reduced exciton density matrix in Fig.~\ref{fig:8_avrdm}. Similar to the case of the seven-site model, there are coherences between site $m=1-3$. Coherences between site $m=8$, however, are only established with sites $m=1$ and 2.
\subsection{Field-Driven Dynamics}
In order to address the effect of initial state preparation, exemplary calculations, explicitly including the radiation-matter interaction, Eq.~(\ref{eq:hfield}), have been performed for the seven-site model. Here, we do not aim at a comparison with experiment. Instead, the focus is on the difference between instantaneous excitation of a local state and the field preparation of a one-exciton eigenstate. Inspecting Fig.~\ref{fig:levels}, one notices that site $m=1$ has the largest amplitude for the one-exciton eigenstate around 400 \cm. However, this eigenstate has also a contribution from site $m=2$. Since the contributions of all other sites are negligible, this situation is well-suited for the present purpose. Thus, we will compare the results of Figs.~\ref{fig:7_pop}-\ref{fig:7_avrdm} with those obtained by explicit excitation of the eigenstate around 400 \cm. To simplify the discussion the summation in Eq.~(\ref{eq:hfield}) is restricted to $m=1,2$. Notice that Fig.~\ref{fig:levels} gives the bare exciton eigenstates only, i.e.\ for the full  Hamiltonian the field will excite an exciton-vibrational wave packet. The field parameters have been chosen such as to give a total excited state population of about 10\% without noticeable contributions of stimulated emission. The pulse is resonant to the bare exciton energy and its spectrum is sufficiently broad (FWHM 89 \cm) to excite vibronic wave packets at the two sites.

Results of the simulations are shown in Fig.~\ref{fig:7_pop_field}. In the lower right panel we give the  amplitude of the field envelope together with the site populations. Comparison with Fig.~\ref{fig:7_pop} reveals that the population dynamics at sites $m=1$ and 2 is rather different. Needless to say that this is not an unexpected result. More interesting is the observation that the population dynamics at the other sites does no differ that much. In other words, under the present excitation conditions, for the population of the site $m=3$, which is attached to the reaction center, it does not really matter whether the system is excited via an ultrashort pulse or by feeding population into site $m=1$. The time-dependence of coherences between the sites in the case of field-driven dynamics shows a less oscillatory behaviour as compared with the instantaneous excitation (see  Fig.~S5 of the Suppl. Mat.\cite{suppl}). 

The similarity between field-free and field-excitation simulations is even more striking for the vibronic and vibrational excitations (compare Figs.~\ref{fig:7_pop} and \ref{fig:7_pop_field}). Indeed, the main difference is a time shift of the excitation due to the finite preparation time of the initial excitation at sites $m=1$ and 2.

\section{Conclusions} 
\label {sec:concl}
In summary, we have investigated the quantum dynamics of coupled excitonic and vibrational DOFs using the ML-MCTDH method for solving high-dimensional Schr\"odinger equations. Thereby, we have extended our previous three-site model~\cite{schulze15_6211} and demonstrated that the consideration of the seven- and eight-site FMO models is indeed computationally feasible. Within these models it was possible to quantify the relative contribution of the major and minor excitation energy transfer pathways. The minor role played by the pathway involving site $m=4$ enabled us to use a hybrid MCTDH/Hartree approach without much deterioration of the resulting final populations. 

The dynamics has been analyzed in terms of the reduced one-exciton density matrix, averaged with respect to the considered time interval. It turned out that in this quantity the effect of EVC on the transfer is reflected in the suppression or enhancement of certain matrix elements as compared to the bare excitonic case. An analysis of the vibrational and vibronic dynamics established that two mechanisms are operative. First, a competition between vibrational motion after Franck-Condon excitation and exciton transfer, which triggers  ground state vibrational dynamics in specific spectral ranges. Second, vibronically-assisted exciton transfer, which yields excitation of a narrow range of vibrational modes in the electronically excited states.

Further, we addressed the issue of initial state preparation. Comparing instantaneous excitations of site $m=1$ and $m=8$ in the seven- and eight-site model, respectively, the scenarios of coherent population oscillations ($m=1$) and quasi-monotonous population decay ($m=8$) have been observed. Finite field excitation has been studied for the seven-site model. Here, it turned out that although there are no coherent oscillations between the populations at sites $m=1$ and 2, the dynamics at the trapping site is rather similar to the case of a sudden excitation at site $m=1$. Moreover, the vibronic and vibrational dynamics is not much affected by the type excitation for the cases studied.
\begin{acknowledgements}
This work was made possible by NPRP grant \#NPRP 7-227-1-034 from the Qatar National Research Fund (a member of Qatar Foundation).
\end{acknowledgements}

%
%
%
%

\end{document}